\documentclass[epj,referee]{svjour}
\usepackage{amssymb}
\usepackage{psfig}
\begin{document}

\title{Gap and pseudogap evolution  within the charge-ordering \\
scenario  for superconducting cuprates}
\author{L. Benfatto, S. Caprara, and  C. Di Castro}

\institute{Dipartimento di Fisica - Universit\`a di Roma ``La Sapienza''
and Istituto Nazionale per la Fisica della Materia - 
Unit\`a di Roma 1,
Piazzale Aldo Moro, 5 - 00185 Roma Italy}

\date{Received: date / Revised version: date}

\abstract{
We describe the spectral properties of underdoped cuprates as
resulting from a momentum-dependent pseudogap in the normal state
spectrum. Such a model accounts, within a BCS approach, for the doping
dependence of the critical temperature and for the two-pa\-rameter
leading-edge shift observed in the cuprates.
By introducing a phenomenological temperature
dependence of the pseudogap, which finds a natural interpretation within
the stripe quantum-critical-point scenario for high-$T_c$ superconductors,
we reproduce also the $T_c-T^*$ bifurcation near optimum doping. Finally,
we briefly discuss the different role of the gap and the pseudogap in 
determining the spectral and thermodynamical properties of the model 
at low temperatures.
} 
\authorrunning{L. Benfatto, S. Caprara, and  C. Di Castro}
\titlerunning{Gap and pseudogap evolution ...}
 \maketitle

\section{Introduction}
The non-Fermi-liquid behaviour of the normal phase of the cuprates has two 
major features: (i) nearby optimum doping the in-plane resistivity is linear
in $T$, signalling the absence of any other energy scale besides the 
temperature; (ii) in the underdoped regime photoemission and tunneling 
experiments show that a pseudogap persists well above the critical temperature 
$T_c$ up to a crossover temperature $T^*$ \cite{timstatt}. While $T_c$ 
increases with 
increasing doping, $T^*$ starts from much higher values and decreases. The two 
temperatures merge around or slightly above optimum doping. Angle-resolved 
photoemission spectroscopy (ARPES) experiments indicate that the pseudogap 
and the superconducting gap have the same momentum dependence across $T_c$, 
which almost resembles a $d_{x^2-y^2}$-wave, namely $\Delta_{\bf k}=
\Delta_0(\phi) \cos (2\phi)$, with $\phi=\arctan (k_y/k_x)$, and that both 
are tied to the underlying Fermi surface \cite{Ding,Norman,Mesot}. In the BCS 
$d$-wave approach $\Delta_0$ is $\phi$-independent and is proportional to 
$T_c$. Here instead $\Delta_0(\phi)$ is angle dependent and the ARPES 
spectra of the pseudogap state evolve smoothly and continuously in the 
superconducting ones across the critical temperature. A smooth evolution is 
also observed in tunneling spectra \cite{Renner}. Moreover, the pseudogap at 
different ${\bf k}$ points opens at different temperatures. At $T^*$ a 
leading-edge shift ($LE$) appears in the ARPES spectra of underdoped Bi2212 
for momenta near the $M\equiv (\pm\pi,0);(0,\pm\pi)$ points of the Brillouin 
zone. The $LE$ is observed as a finite minimum distance of the quasiparticle 
peak from the Fermi level in the non superconducting state, and leaves 
disconnected arcs of Fermi surface. When the temperature is lowered the $LE$ 
regions around the $M$ points enlarge and the arcs of Fermi surface reduce 
and shrink towards the nodal points of the corresponding $d$-wave 
superconducting gap below $T_c$. At the same time the doping dependence of 
the momentum structure of the $LE$ is not trivial. As the doping is 
increased, the zero temperature $LE$ at the $M$ points, $\Delta_0(0)$, 
remains constant or decreases \cite{Ding,Norman,Mesot,deut}, while the 
$LE$ around the nodal points $\Delta_0(\pi/4)$ seems to increase
\cite{Mesot,Xiang,deut} and follow the
rising of the critical temperature. Penetration depth measurements of the 
superfluid density $\rho_s(T)$ at low temperature probe the low-energy 
excitations around the nodal points in a $d$-wave superconductor, and 
therefore $\Delta_0(\pi/4)$. The correspondence between ARPES measurements
of $\Delta_0(\pi/4)$ and the slope of $\rho_s(T)$ at $T=0$ is
made however more involved by the presence of the Landau renormalization
factors \cite{Mesot}.

Many theoretical models have been proposed to obtain a non-Fermi-liquid 
behaviour and to describe a pseudogap state. A firm result is however that  
above one dimension the Landau Fermi-liquid theory is generically stable
and a strongly singular effective potential is required to disrupt it
\cite{metz}. This result, together 
with the above phenomenology, suggests that a consistent description of the 
cuprates requires a strong momentum-, doping-, and temperature-dependent 
effective interaction. This interaction  should  affect the states near the 
$M$ points of 
the Brillouin zone more effectively than along the diagonals and should have 
the temperature as the only energy scale around optimum doping. It was
shown that in 
strongly correlated systems in the presence of additional attractive 
interactions (e.g. the Hubbard-Holstein model) and of long-range Coulomb 
forces, the exchange of quasi-critical charge (and spin) fluctuations 
provides such an effective electron-electron interaction both in the 
particle-particle and in the particle-hole channel \cite{prl95}. 
Non-Fermi-liquid behaviour 
and strong pairing mechanism have in this way a common origin. These 
fluctuations arise near a finite-temperature instability line 
$T_{CDW}(\delta)$ for charge-density wave or stripe-phase formation, which  
ends in a quantum critical point (QCP) at $T=0$ and $\delta=\delta_c$ 
near optimum doping \cite{prl95,prb96}. As shown in Ref. \cite{prl95}, the 
effective 
electron-electron interaction near the charge instability has the form
\begin{equation}
V_{eff}({\bf q},\omega)\simeq \tilde{U}-\frac{V}{\kappa^2+|{\bf q}-
{\bf q}_c|^2-{\rm i}\gamma\omega},
\label{vq}
\end{equation}
both in the particle-hole and particle-particle channels. Here
${\bf q}\equiv (q_x,q_y)$ and $\omega$ are the exchanged momenta and
frequencies in the quasiparticle scattering, $\tilde{U}$ is a residual 
repulsion, $V$ is the strength of the attractive effective potential, 
${\bf q}_c$ is the critical wave-vector related to the charge ordering 
periodicity ($q_c=2\pi/\lambda_c$). For physically relevant values of the 
parameters of the Hubbard-Holstein model ${\bf q}_c$ turns out to be 
$(\pm 0.28,\pm 0.86)$ or equivalently $(\pm 0.86,\pm 0.28)$ \cite{prl95}. 
In this case, therefore, ${\bf q}_c$ connects the
two branches of the FS around the $M$ points and strongly affects these states.
The mass term $\kappa^2= \xi_c^{-2}$ is the inverse square of the correlation 
length of the charge order and provides a measure of the distance from 
criticality. This is given by $\delta-\delta_c$ in the overdoped region, by 
$T$ in the quantum critical region around $\delta_c$ and by 
$T-T_{CDW}(\delta)$ in the underdoped region, where $T_{CDW}(\delta)$ sets in 
a new doping-dependent energy scale closely followed by $T^*(\delta)$. The 
characteristic time scale of the critical fluctuations is $\gamma$. The 
presence of a weak momentum-independent repulsion $\tilde{U}$ together
with a strong attraction of the order of $-V/\kappa^2$ in the 
particle-particle channel (cfr. Eq. (\ref{vq}) favors $d$-wave 
superconductivity approaching optimum doping from the overdoped regime, 
within direct BCS calculations \cite{prb96}. In the underdoped regime 
we expect that precursor effects of charge ordering are relevant to the
pseudogap formation and extend up to a temperature $T_0(\delta)$, (the mean 
field temperature for CDW formation) higher then $T_{CDW}\sim T^*$. 

The two limiting cases when these precursors dominate the pseudogap 
formation in a single channel only (either particle-particle or 
particle-hole channel) are simpler to 
analyze and each of them shows relevant aspects of the physics of the 
cuprates. The interplay of the two channels is an open problem still under 
investigation. 

A first possibility is that the pseudogap opens due to 
incoherent paring in the particle-particle channel, leading to a state 
where Cooper pairs around the $M$ points are formed at $T^*\gtrsim T_{CDW}$ 
with strong long-wavelength fluctuations. Phase coherence, which 
characterizes a real superconducting state, is established at a lower 
temperature $T_c$, by 
coupling to the stiffness of the pairing near the nodal points \cite{2gap}.

In this paper we elaborate the other possibility, that the transition to the 
superconducting state takes place in the presence of a normal-state pseudogap 
parameter $\Delta_p$ resulting from interactions in the particle-hole 
channel. The issue arises
of the interplay between the preformed pseudogap in the p-h channel and the
additional pairing in the p-p channel. 
Having included most of the anomalous effects in the pseudogap formation, we 
determine $T_c$ via the BCS approach for the pairing in the p-p channel.
Our model 
originates as a simple schematization of a system interacting via the 
singular effective interaction (\ref{vq}) and is inspired to a similar 
model proposed by Nozi\`eres and Pistolesi \cite{NP}, with the inclusion of
some specific aspects of the phenomenology of the cuprates. In Section 2 
we discuss the model for the normal-state spectrum in presence of a
pseudogap which has a $d$-wave form with amplitude $\Delta_p$. Assuming at
the beginning a costant $\Delta_p$, we discuss in Section 3 the general
properties of our model, devoting a particular attention to the doping and/or
temperature dependence of $T_c$, of the $LE$ and of the superfluid
density. In Section 4 we introduce a modulation for $\Delta_p$, to take 
care of the $\delta$-dependence of the new energy scale set by the 
$T_{CDW}(\delta)$ in the underdoped case. We assume that the pseudogap
opens nearby a mean-field temperature $T_0(\delta)$ for the onset of $CDW$. 
$T_0(\delta)$ should follow the doping dependence of 
$T_{CDW}(\delta)$ in the underdoped regime and produce a variation in the density of states, as 
revealed by NMR and resistivity measurements on several compounds
\cite{timstatt}. By a suitable fitting of $T_0(\delta)$, we give at the end a
phenomenological description of the phase diagram of the cuprates, together
with some physical quantities like the superfluid density, the specific
heat and the leading edge.

\section{The model}
Within the above scenario, we describe the  pseudogap in the normal state 
by means of a simplified model where a ${\bf k}$-dependent separation is
present between a valence band and a conduction band, as a result of a 
${\bf k}$-dependent effective interaction in the particle-hole channel.   
Differently from Ref. \cite{NP}, we 
adopt a lattice electron model and assume that the pseudogap vanishes at 
some points of the Brillouin zone. Being interested on very qualitative 
aspects of the evolution of the pseudogap state, we shall mainly concentrate 
on a two-dimensional system related to the CuO$_2$ planes, the third 
dimension being relevant to establish the nature of the true transition and to 
cut off the corresponding fluctuations. Accordingly, we model the 
normal-state spectrum as
\begin{equation}
\xi_{\eta \bf k}=-\mu+\eta \sqrt{\epsilon_{\bf k}^2+\Delta_p^2
\gamma_{\bf k}^2}, 
\label{eqxi}
\end{equation}
where $\eta=+1(-1)$ in the conduction (valence) band, $\epsilon_{\bf k}=
-2t(\cos k_x+\cos k_y)$ is the tight-binding  dispersion law in the 
conventional metallic state (i.e. at $\Delta_p=0$), $\Delta_p$ is the
pseudogap parameter, and $\mu$ is the chemical potential. $\Delta_p$
governs together with the chemical potential the $LE$ opening at the $M$
points. The modulation 
of the pseudogap is given by the absolute value of $\gamma_{\bf k}=(\cos k_x
-\cos k_y)/2$, which vanishes along the diagonals of the Brillouin zone. 
Since $\epsilon_{\bf k}$ vanishes along the lines $k_y=\pm(\pi\pm k_x)$,
the two bands merge at the points ${\bf k}_P=(\pm \pi/2,\pm \pi/2)$.
In the undoped system the $LE$ has the ${\bf k}$-modulation given by the
factor $\gamma_{\bf k}$.  As the system is doped with respect to 
half-filling, a Fermi surface appears  in the form of small pockets around 
the ${\bf k}_P$ points. The $LE$, which  vanishes along the Fermi surface, 
evolves thus continuously with respect to the undoped case. This is in 
contrast with the case of a ${\bf k}$-independent pseudogap $\Delta_p$, 
which leads, in the lattice model, to a large Fermi surface in the doped 
system, with the chemical potential and the $LE$ which evolve
discontinuously with respect to half-filling. 

\begin{figure}[h]
\psfig{figure=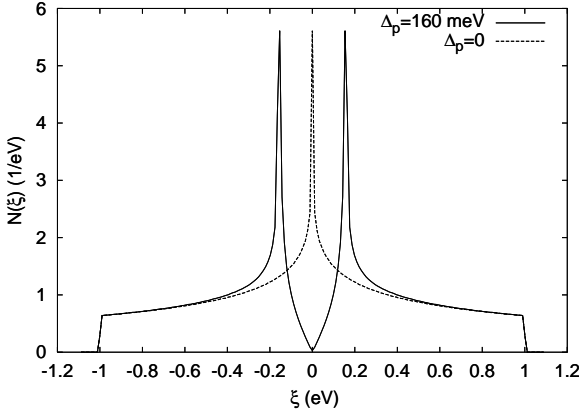,width=8cm,angle=-90}
\caption{\footnotesize Density of states at half-filling in the presence
and in the absence of pseudogap for a hopping parameter $t=0.25$
eV. Observe in the case $\Delta_p=160$ meV the splitting of 
the van Hove singularity which gives rise to two peaks at
energies $\xi \approx \pm \Delta_p$, with 
$\Delta_p\ll 2t$.}
\label{figdos}
\end{figure}
We are aware of the 
fact that the complicated band structure of the cuprates is not simply fitted 
by the form (\ref{eqxi}). For instance, the band structure can be improved by 
extending the tight-binding expression beyond nearest neighbours. However, 
this does not represent a severe limitation to our approach, the 
generalization being straightforward. Moreover, the presence of the 
pseudogap leads, within our model, to the appearance of small pockets in
the weakly doped system. This issue is experimentally controversial, and it 
seems definitely confirmed that the two branches of each pocket, even when 
observed \cite{shadow}, are not equivalent \cite{boris}. However, the 
presence of pockets around the ${\bf k}_P$ points, which result from the 
oversimplification of our description, is not essential to reproduce the 
finite $LE$ observed in the single-particle ARPES spectra near the $M$ 
points, and does not play any significant role in the forthcoming analysis. 
Therefore, we shall not further discuss this aspect.

The density of states corresponding to the band structure (\ref{eqxi})
vanishes only at $\xi=-\mu$, and is finite elsewhere, as shown in Fig. 1.
The van Hove singularity which exists at $\xi=-\mu$ for $\Delta_p=0$ is 
thus split into two singularities  separated by 
$4t\Delta_p/\sqrt{(\Delta_p/2)^2+4t^2}$. For $\Delta_p \ll 2t$, as we shall
assume in most of our calculation, the energy range where the density of
states in suppressed is of order $2\Delta_p$.

Having included most of the effects of the anisotropic potential in the
${\bf k}$-dependence of the pseudogap $\Delta_p |\gamma_{\bf k}|$, for simplicity
we assume that the onset of superconductivity is produced within a $BCS$
approach by a constant pairing interaction among the carriers in 
the Cooper channel, apart again from a $d$-wave modulation. The
relevance of the superconductive fluctuations of the phase of the order 
parameter will be discussed below, in parallel with the analysis of the 
properties of the superfluid density.

We introduce therefore an intraband interaction term between time-reversed 
states 
\begin{eqnarray}
H_I=-V \Omega \sum_{\eta} & &\int \frac{d^2 k}{(2\pi)^2} \gamma_{\bf
k} \gamma_{\bf k'} c^+_{\eta, \bf k \uparrow}c^+_{\eta, \bf
-k \downarrow} \times \nonumber\\
& &\int \frac{d^2 k'}{(2\pi)^2}
c_{\eta, \bf -k' \downarrow}c_{\eta, \bf k' \uparrow},
\label{eqhi}
\end{eqnarray}
where $V$ is the pairing strength, $\Omega$ is the volume of the system, 
$c^+_{\eta,\bf k,\sigma}$ is the creation operator of the electrons in the 
$\eta$-band, and the factors $\gamma_{\bf k}$ make the $d$-wave symmetry 
explicit. Introducing the order parameters
$
\Delta_{\eta}=V \langle\gamma_{\bf k} c^+_{\eta, \bf k \uparrow}c^+_{\eta, \bf
-k \downarrow}\rangle
$
we obtain the self-consistency equations
\begin{eqnarray}
\Delta_\eta&=&\frac{V}{2} \int \frac{d^2 k}{(2\pi)^2}
\frac{\gamma_{\bf k}^2}
{E_{\eta \bf k}}\tanh \left(\frac{\beta E_{\eta \bf k}}{2}\right) \Delta_\eta
\qquad (\eta=\pm 1),
\label{eq3}\\
\delta&=&1-\frac{1}{2}\sum_{\eta} \int \frac{d^2 k}{(2\pi)^2}
\left[1- \frac{\xi_{\eta \bf k}}{E_{\eta \bf k}}
\tanh \left(\frac{\beta E_{\eta \bf k}}{2}\right) \right],
\label{eq4}
\end{eqnarray}
where $E_{\eta \bf k}=\sqrt{\xi^2_{\eta \bf k}+\Delta_\eta^2
\gamma_{\bf k}^2}$ is the quasiparticle energy in the superconducting state. 
The last equation fixes the chemical potential $\mu$ for any given doping 
$\delta$ with respect to half-filling. 
As most of the cuprates become superconducting by doping with holes, 
we study the hole-doped regime $\delta>0$, where the chemical potential at 
$T=0$, in the absence of pairing, falls in the valence band. Therefore, at 
weak-coupling, we find the solution $\Delta_{\eta=-1}=\Delta$,  
$\Delta_{\eta=+1}=0$  for Eq. (\ref{eq3}). We point out that, due to the 
assumed band structure (\ref{eqxi}), the spectrum is particle-hole 
symmetric, and doping with electrons leads to the same self-consistent 
solution, provided the role of the two bands is interchanged. 

Adding to the 
Hamiltonian (\ref{eqhi}) an interband interaction term, as discussed in
Ref. \cite{NP}, induces the $\Delta_{\eta=+1}$ to be different from
zero even at weak-coupling. The two order parameters are generically 
different, contrary to the assumption of Ref. \cite{NP}. Indeed, the 
solution with one order parameter ($\Delta_{\eta=+1}=\Delta_{\eta=-1}$) 
has  a higher free energy, which may become equal under very specific
conditions. 

\section{General properties of the model}

In this section we shall analyze the main outcomes of our model and
their dependence on doping $\delta$ and on the temperature $T$. The 
parameters $\Delta_p$ and $V$ at the beginning will be assumed to be
temperature and doping independent, and not related to each other. This 
preliminary study allows us to point out the consequences of the assumed 
${\bf k}$-dependent normal-state pseudogap, and to single out the regime of 
parameters suitable for a description of the phenomenology of the cuprates. 

By solving the Eqs. (\ref{eq3})-(\ref{eq4}) we obtain the temperature 
and doping dependence of the superconducting gap $\Delta$ and of the chemical 
potential $\mu$, and we determine the phase diagram of the model in different 
physical regimes.

Before analyzing the two regimes of weak or strong coupling at various
doping, we investigate the existence of a finite strength 
$V_c$ of the interaction to have superconductivity in the undoped system,
due to the fact that at half-filling the density of states vanishes 
at the Fermi energy. 
In our model, due to the $d$-wave nature of the preformed gap, 
the value of $V_c$ is less than that found in  Ref. \cite{NP} in presence
of a constant gap. The equation that defines $V_c$ at $T=0$, $\delta=0$ 
as a function of the pseudogap amplitude $\Delta_p$ is
\begin{equation}
\frac{2t}{V_c}=\frac{1}{2}\int \frac{d^2 k}{(2\pi)^2}\frac{\gamma_{\bf
k}^2}{\sqrt{(\cos k_x+\cos k_y)^2+(\Delta_p/2t)^2\gamma_{\bf k}^2}}
\label{eqvc}
\end{equation}
so that $V_c\approx \pi^2 \Delta_p/2$, for $\Delta_p\gg 2t$, and 
$V_c\sim 2t/\ln(4t/\Delta_p)$ for $\Delta_p \ll 2t$. The former limiting case 
is however not realistic: henceforth we shall definitely assume $\Delta_p\ll
2t$. Moreover, in the present scenario, the pseudogap is associated 
with scattering in the particle-hole channel. Pairing must therefore be 
absent at half-filling and we assume as the relevant regime for the
cuprates $V<V_c$.

{\em Critical temperature} Within the BCS approach, the variation of the 
critical temperature with doping 
is controlled by the density of states at the Fermi level. At $\Delta_p=0$, 
we recover the spectrum of a normal metal with nearest-neighbours hopping: at 
half-filling the Fermi energy is at the van Hove singularity, and $T_c$ is 
maximum. By increasing doping, the density of states at the Fermi energy 
decreases, and so does $T_c$. This behaviour is excluded by the experiments. 
On the contrary, when $\Delta_p>0$, the critical temperature is zero at 
half-filling (if $V<V_c$). At fixed finite doping, the critical temperature 
increases with increasing the pseudogap parameter $\Delta_p$, reaches a 
maximum, and then decreases. By doping, the critical temperature follows the 
evolution of the density of states at the Fermi energy, reaching a maximum 
when the Fermi energy passes through the boundary of the pseudogap region in 
the density of states, i.e. at the doping $\delta_{opt}$ such that 
$\mu \approx -\Delta_p$. For $\delta>\delta_{opt}$, $T_c$ decreases with 
increasing doping. The resulting bell-shaped $T_c$ vs $\delta$ curve captures 
the main features of the experimental results, as it will be more
accurately presented in Section (4) with a doping dependent $\Delta_p$.

{\em The leading edge} 
The quasiparticle spectrum, both in the normal and in the superconducting 
state, is characterized by a leading-edge shift $LE$, which is defined as
$LE(\phi)=\min_{\eta,k}E_{\eta\bf k}$, where ${\bf k}= k(\cos \phi,\sin
\phi)$ \footnote {Within our mean-field treatment the excitations have a
well defined energy and this quantity should be more correctly called
``one-particle excitation gap''. However, by using the term $LE$ we want to
put more emphasis on the distribution of spectral weight than on its 
coherent character, which is an artifact of our description of the cuprates
in terms of a semiconducting band structure.}.
Due to the point symmetries of the system our analysis can be limited 
to the quadrant $0\leq \phi \leq 45^\circ$. In the strong coupling regime 
the superconducting gap $\Delta \gg \Delta_p$, so the
doping variation of $\Delta$ dominates the evolution of the zero-temperature
$LE$ at the $M$ points, which increases approaching optimum doping, following 
$T_c$, in contrast with the behaviour observed in the experiments in the 
underdoped regime. In the weak-coupling regime instead the superconducting 
gap is much smaller than the pseudogap. For $\delta < \delta_{opt}$ the
$LE$ of the superconducting state is controlled by two independent 
parameters (see Fig. (2)), whose existence is also
suggested by the combination of ARPES, penetration-depth and Andreev
experiments \cite{Ding,Norman,Mesot,Xiang,deut}. 
The amplitude at the $M$ points ($LE_M$) is given 
essentially by the value of the $LE$ in the normal state, i.e.
\begin{equation}
LE_M =  \frac{2t\Delta_p}{\sqrt{(\Delta_p/2)^2+4t^2}}-|\mu| \approx
\Delta_p-|\mu| \qquad (\Delta_p\ll 2t) 
\label{eqle}
\end{equation}
independent of $T_c$, whereas the slope at the nodes is 
$v_{\Delta}=\Delta_0(\phi=\pi/4)\propto \Delta \propto T_c$, in agreement
with the experimental finding \cite{Mesot,Xiang,deut}. Note instead that $LE_M\approx 
\Delta$ for $\Delta_p=0$.

\begin{figure}[h]
\psfig{figure=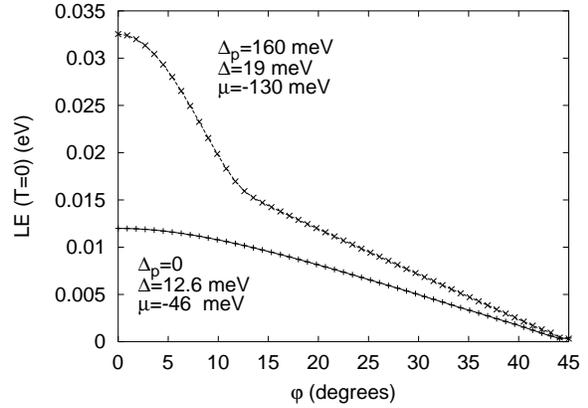,width=8cm,angle=-90}
\caption{\footnotesize Zero-temperature $LE$ vs $\phi$ for $t=250$ meV, 
$\delta=0.1$, in the weak-coupling regime $V=87$ meV, in the presence
($\times$)  and in the absence ($+$) of the normal-state pseudogap 
$\Delta_p$. Observe, for $\Delta_p=160$ meV,  the crossover around 
$\phi \approx 12^\circ$ to a regime dominated by the normal state pseudogap.}
\label{fig_1}
\end{figure}

The variation of the chemical potential accounts now for 
the main dependence of the $LE$ on doping. By increasing doping the 
normal-state $LE_M$, controlled by the pseudogap parameter 
$\Delta_p$ according to Eq. (\ref{eqle}),  decreases and vanishes at a
doping $\delta_c$. If $\Delta_p$ is doping independent (as we are assuming 
in this preliminary discussion), the $\delta_c$ for which $LE_M=0$
coincides with the doping $\delta_{opt}$ for the highest $T_c$. 
At $\delta>\delta_c$ the $LE_M$ in the superconducting state is then
given only by the superconducting order parameter $\Delta$, as shown in
Fig. 3. At fixed  
doping, the temperature variation of the $LE$ depends on the closing of the 
superconducting gap at $T_c$, and eventually on the temperature dependence 
of $\Delta_p$, if any (see Fig. 4).

\begin{figure}[h]
\psfig{figure=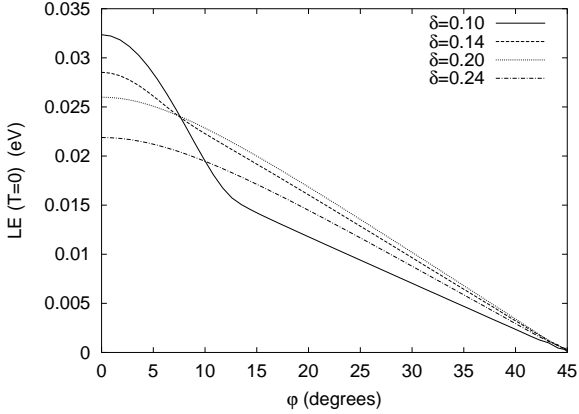,width=8cm,angle=-90}
\caption{\footnotesize 
Doping dependence of the zero temperature $LE$ for $t=250$ meV, 
$\Delta_p=160$ meV and $V=87$ meV. The $LE_M=LE(\phi=0)$ decreases with
increasing doping, due to the reduction of the normal-state $LE_M$ given by
Eq. (\ref{eqle}). At $\delta>\delta_{opt} \approx 0.2$ the normal state 
$LE$ closes and $LE(\phi, T=0)$ is that of a standard $d$-wave 
superconductor (see also Fig. 2). Instead, the slope of the $LE$ 
at the node $v_{\Delta}\propto \Delta$ increases before reaching
$\delta_{opt}$ and then decreases. The critical temperature (proportional
to $\Delta(T=0)$ in the BCS approach) follows the same doping dependence.}
\end{figure}

\begin{figure}[h]
\psfig{figure=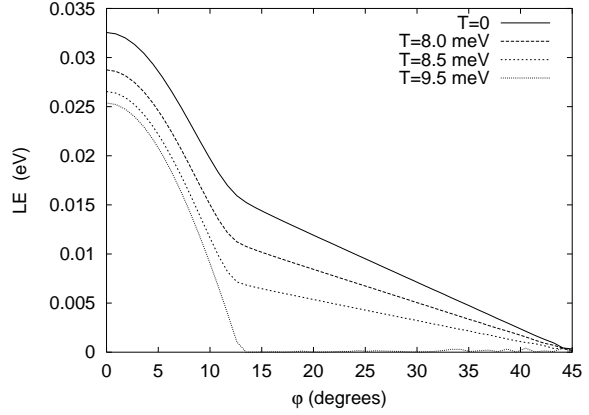,width=8cm,angle=-90}
\caption{\footnotesize 
Temperature dependence of the $LE$ for the same set of parameters of Fig. 3 
(increasing temperature from top to bottom). At $T>T_c= 9$ meV the 
superconducting gap closes, leaving a finite $LE_M\approx \Delta_p-|\mu|$.}
\end{figure}

{\em The superfluid density} 
In the superconducting phase the thermodynamic properties at low
temperature are essentially determined by the quasiparticles near the
nodes, i.e. by the superconducting gap $\Delta$, whereas the pseudogap 
amplitude $\Delta_p$ plays no role. The superfluid density for instance can
be evaluated according to the simple BCS formula:
\begin{eqnarray}
\rho_{s}=
\frac{1}{2}\sum_{\eta,l=x,y} \int \frac{d^2 k}{(2\pi)^2}& & \left\{
\frac{\partial^2 \xi_{\eta \bf k}}
{\partial k_l^2} \left(1- \frac{\xi_{\eta \bf k}}{E_{\eta \bf k}}
\tanh\frac{\beta E_{\eta \bf k}}{2}\right)\right .+\nonumber\\
& &\left . 2\left(\frac{\partial \xi_{\bf k}}{\partial k_l}\right)^2 
\frac{\partial f(E_{\eta \bf k})}{\partial E_{\eta \bf k}}\right\},
\label{eqrho}
\end{eqnarray}
where $f(x)$ is the Fermi function. $\rho_s$ decreases linearly in $T$ at 
low temperature, as expected in a $d$-wave superconductor, with a slope 
determined only by the superconducting gap. For the assumed band dispersion 
(\ref{eqxi}) it can be shown that
\begin{equation}
\rho_s(T)-\rho_s(0)\simeq  -\frac{T}{\Delta}\frac{16t\ln 2}{\pi}.
\label{eqsl}
\end{equation}
The slope of the superfluid density at low temperature, $\alpha=d\rho_s
/dT(T=0)$ is estimated in Ref. \cite{Mesot} by means of direct ARPES 
measurements of the
slope $v_{\Delta}$ of the superconducting gap near the node. Because the
observed $v_{\Delta}$ decreases in underdoped region (contrary to the
gap at the $M$ points), the doping dependence of $\alpha \propto
1/v_{\Delta}$ is found to increase by decreasing doping with respect to 
its optimum value. This behaviour is confirmed by our model: the slope of 
$d\rho_s/dT$ given in Eq. (\ref{eqsl}) is controlled by the superconducting 
gap $\Delta$, which follows the doping dependence of $T_c$ in the 
underdoped regime. However, as observed in Ref. \cite{Mesot}, the general 
trend of $\alpha(\delta)$, obtained by direct measurements of penetration 
depth, seems to be opposite, giving a decrease  in the underdoped compounds 
of about $40 \%$ respect to its value at optimum doping. The 
correspondence  between ARPES estimates and experimental data on
$\alpha(\delta)$ requires the inclusion of doping dependent Landau
renormalization factors, as discussed in Ref. \cite{Mesot} using the
expression for $\rho_s$ with the inclusion of quasiparticles interaction
(see also Ref. \cite{millis}). The presence of a doping dependent Landau
factor is plausible within our scenario if we attribute the origin of
quasiparticle scattering to quasi-critical charge and spin fluctuations as
in Eq. (\ref{vq}). The same holds for a correspondence between 
the slope of the
density of states $N(\xi)$ near $\xi=0$, where $N(\xi)\approx \eta \xi$
with $\eta \propto 1/v_\Delta$, and the values of $\eta$ extracted by
specific heat measurements \cite{PX99}. 

As the temperature is increased, the system crosses 
over to a regime of higher energy excitations, which sample regions where 
$LE>\Delta$, and the slope is reduced. This can be seen  in Fig. 5,
where the temperature variation of $\rho_s(T)$ in weak and
strong coupling is reported. The variation of the slope with respect to its 
low temperature value is  less pronounced in the strong coupling case, where 
almost the same parameter controls the $LE_M$ and the slope of the
superconducting gap at the node.

\begin{figure}[h]
\psfig{figure=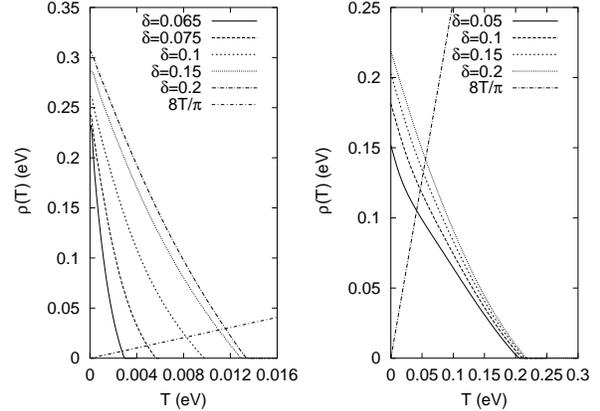,width=8cm,angle=-90}
\caption{\footnotesize $\rho_s(T)$ at different doping for $\Delta_p=0.16$
eV, $t=0.25$ eV in weak coupling ($V=0.09$ eV, left) and strong coupling
($V=0.9$ eV, right). The temperature at which the numerical curves intersect
the line $8T/\pi$ is the $T_{KT}$, as explained in the text. Phase 
fluctuations are important between $T_{KT}$
and $T_{BCS}$.}
\label{fig_2}
\end{figure}

Finally, we consider the doping dependence of the superfluid density. 
The variation of the $\rho_s(0)$ with doping is related 
to the variation  of the number of carriers with doping: in a lattice model 
it does not increases monotonically, but reaches a maximum at an intermediate 
$0<\delta<1$, and then decreases. Nevertheless, the doping for the maximum 
$\rho_s(0)$ does not coincides necessarily with $\delta_{opt}$. 

The temperature dependence of the superfluid density allows us to give an 
estimate of the range of temperatures in which the phase fluctuations of the 
superconducting order parameter are relevant. Indeed, the critical 
temperature for phase coherence in a two-dimensional system, $T_{KT}$, is 
related to the phase stiffness $\rho_s$ via the Kosterlitz-Thouless relation 
$\rho_s=8T_{KT}/\pi$. Assuming the BCS temperature dependence for 
$\rho_s(T)$, the above relation becomes a self-consistency condition 
$\rho_s(T_{KT})=8T_{KT}/\pi$, and the phase fluctuations play an important 
role in the range of temperatures between  $T_{KT}$ and $T_{BCS}$. As it is
naturally, in our 
model this region is very narrow in weak coupling, and becomes larger and 
larger as the strength of the pairing interaction increases. Indeed while 
$\rho_s(0)$ mainly depends on the doping $\delta$, $T_{BCS}$ rapidly 
increases with increasing the coupling, and becomes much larger than $T_{KT}$.
In Ref. \cite{NP}, due to the discontinuity of the density of states at the
band edges, at low doping a regime $T_c\gg T_{KT}$ is found. In such a case
the transition is expected to be Kosterlitz-Thouless. In our model the
density of states at the Fermi level vanishes smoothly and continuosly as
half-filling is approached. Therefore, in the weak-coupling limit, the 
critical temperature increases slowly with doping, and the
Kosterlitz-Thouless regime is never found. Consistently with our
assumptions within this model, in the regime of interest for cuprates, 
i.e. at weak coupling, the superconducting transition is essentially 
BCS-like, even in the underdoped region. The main part of the non classical 
behaviour is enforced by the presence of $\Delta_p$.

Phase fluctuations effects have also been invoked to explain the linear-$T$
behaviour of $\rho_s(T)$ \cite{EK}. Within a model with pseudogap formation 
in the p-p channel \cite{2gap} these effects deserve a careful analysis.

\section{Conclusions and discussion}

In conclusion we want to summarize the previous results in closer
connection with the phenomenology of the cuprates.
The discussion of the previous section has given some preliminary
indications on the set of parameters suitable to reproduce the general
shape of the  phase diagram observed in the cuprates. Nevertheless, to be 
consistent with the experiments, we  have to introduce the temperature and 
doping dependence of the pseudogap parameter $\Delta_p(T,\delta)$.

\begin{figure}[h]
\psfig{figure=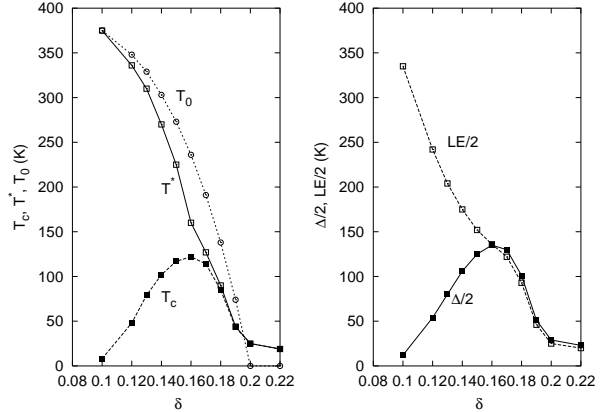,width=8cm,angle=-90}
\caption{\footnotesize  Left panel: doping dependence of $T^*$, 
$T_0$ and of the critical temperature $T_c$. Right panel: doping
dependence of the superconducting gap $\Delta(T=0)$ and of the zero 
temperature $LE_M$. The set of parameters used is $t=250$ meV, $V=85$ meV, 
$c=7$, $\delta_0=0.2$, and we assumed $T_0(\delta)=40
[1-(\delta/\delta_0)^4]$ meV. The value of $T^*(\delta=0)=T_0(\delta=0)$
has been extrapolated from Ref. \cite{Ding}.}
\label{fig_3}
\end{figure}

The mechanism for pseudogap formation has not been the main issue in this 
paper, which focuses on the properties of superconductivity arising in the
pseudogap state. Nevertheless, according to the discussion of Sec. 1, 
we can imagine 
that $\Delta_p(T,\delta) \gamma_{\bf k}$ schematizes the whole complication 
of the strong scattering due to quasicritical charge fluctuations in the 
p-h channel, which arises in the stripe-QCP scenario \cite{prl95}. We assume 
that the doping dependence of $\Delta_p$ follows the doping variation of
the temperature $T_0$ $(T_0>T^*)$ at which a first variation in the density 
of states occurs, as revealed by NMR and resistivity measurements
\cite{timstatt}. We adopt the simple relation:

\begin{equation}
\Delta_p(T,\delta)= c T_0(\delta) g\left(T/T_0(\delta)\right),
\label{eqdd}
\end{equation}
where $g(1)=0$, $g(0)=1$. $g(x)$ interpolates smoothly between these 
two limits \footnote{Here we assume $g(x)=(1-x^4/3)\sqrt{1-x^4}$, which
reproduces a mean-field-like  behaviour near $T=0$ and
$T_0$. However, the specific form of $g(x)$ does not play a crucial role in
determining the general shape of the phase diagram.} and $c$ is a constant
which we use as a fitting parameter. Since 
the identification of $T_0(\delta)$ via NMR 
and resistivity measurements leaves large uncertainty, we decide to refer to
the ARPES experiments, which however only provide the temperature $T^*$ of 
$LE$ closure. We proceed in the following way. In the heavily underdoped 
regime the
main temperature dependence of the $LE_M$ (at fixed doping) is due to the
decreasing of $\Delta_p$ with increasing the temperature, rather than to
the temperature variation of the chemical potential. Then $T^*$ and $T_0$ 
coincide at low doping in our model, whereas the experimental data seem to
indicate that $T_0$ stays larger then $T^*$. This is an artifact of our 
approach, which is not devoted to establish the connection between the 
pseudogap state and the Mott insulator and/or the antiferromagnetic phase. 
This is a relevant  open problem still under investigation. 
Meanwhile we put $T_0(\delta=0)=T^*(\delta=0)$ and 
extrapolating to $\delta=0$ the $T^*(\delta)$ dependence reported in Ref. 
\cite{Ding} for Bi2212 we obtain $T_0(\delta=0)=40$ meV. 
The constant $c$ and the coupling $V$ are adjusted to fix the values of the 
$LE$ in the underdoped regime and the maximum $T_c$. The temperature $T_0$
is assumed to decrease with increasing doping and to vanish at 
$\delta_0=0.2$, slightly above the optimum doping, as expected in the QCP 
scenario \cite{prl95}. Finally, we interpolate between $\delta=0$ and
$\delta_0$ with the expression  $T_0(\delta)=40 
[1-(\delta/\delta_0)^4]$ meV, which allows us to reproduce the shape of the
curve $T^*(\delta)$ observed in the underdoped regime.

The resulting phase diagram is shown in Fig. 6. Even within a simplified
description of the doping and temperature dependence of $\Delta_p$, we
recover the bifurcation of $T^*$ and $T_c$  observed in the
underdoped regime, while they merge around optimum doping. $T_c$
follows the typical bell-shaped curve as a function of $\delta$.
Our approach is still lacking of the temperature and doping dependence of 
the superconducting coupling, which plays an important role in determining 
the $T_c(\delta)$ variation around the QCP, as it has been shown elsewhere 
\cite{prl95,prb96}. Even though the introduction of this further complication 
would definitively improve the agreement with the experimental data, with a
faster decay of $T_c$ at high doping, it would not change the main features 
of the pseudogap state obtained here.
We reproduce the variation of the zero temperature $LE_M$ 
in all the phase diagram: according to the general discussion of the
previous section, the $LE_M$ is determined by the normal state pseudogap in 
the underdoped region, and by the superconducting gap in the overdoped
regime. As a consequence, in the underdoped regime the $LE_M$ is
uncorrelated to the characteristic energy scale of low temperature 
quasiparticle excitations, which probe the value of the superconducting
order parameter around the nodal points. Having now a doping and
temperature dependent $\Delta_p$, 
the doping for the maximum $T_c$ does not coincide anymore with the doping 
at which the normal state leading edge closes, as we had in the presence of 
a constant $\Delta_p$. Indeed,  approaching optimum doping
$\Delta_p$ itself closes at a lower temperature, and as a consequence the
normal state $LE_M$ closes faster than it would be under the only effect of 
the variation of the chemical potential with doping.

\begin{figure}[h]
\psfig{figure=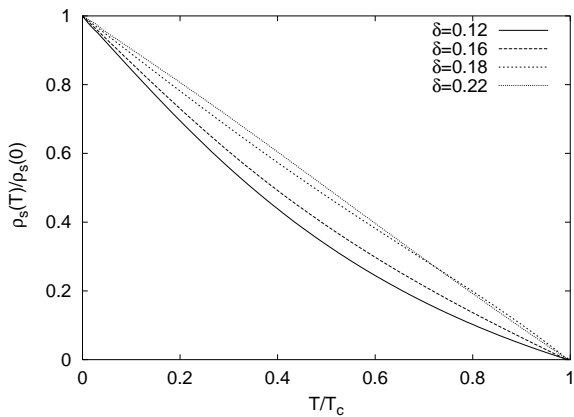,width=8cm,angle=-90}
\caption{\footnotesize  Doping dependence of the normalized superfluid
density $\rho_s(T)/\rho_s(0)$ as a function of $T/T_c$. We used the same
set of parameters of Fig. 4. According to the phase diagram of Fig. 4, the
optimum doping is at $\delta_{opt}\approx 0.16$. Observe that in the 
overdoped regime $\delta>\delta_{opt}$ the superfluid density has the
standard $d$-wave mean field behaviour.}
\label{fig_5}
\end{figure}

Finally, we report in Fig. 7 the normalized superfluid density
$\rho_s(T)/\rho_s(0)$ as a function of $T/T_c$ at various doping. In the
overdoped regime, where the pseudogap closes, we recover the standard
$d$-wave mean field result, which seems in a good agreement with
experimental data \cite{PX99} (see the curve for $\delta=0.22$ in Fig. 7). 
However, in the underdoped region the
dependence on doping of $d\rho_s/dT(T=0)$ does not agree with the
experiments as already discussed in Section (3), where the possible
improvement due to Landau factors has been adressed. We observe here
another discrepancy: the slope of $\rho_s(T)$ near $T_c$ in our model
reduces by increasing the temperature contrary to the experimental findings
\cite{PX99}. As already indicated in Section (3), within our model the
superconducting fluctuations are only important in the short region between
$T_{KT}$ and $T_{BCS}$. The inclusion of these fluctuations, by reducing
$\rho_s$ near $T_c$, would at least partially take care of this
discrepancy. 

Even though we did not reach yet the point to exploit fully the momentum
and doping dependence of a quasi-singular effective interaction among
quasiparticles arising nearby an instability in the p-p and p-h channels,
altogether within the present simplified approach we produce a behaviour of
$T_c(\delta)$, of $T^*(\delta)$ and of the $LE_M(0,\delta)$ in reasonable
agreement with the experimental findings. 

\vspace{1.5cm}
\noindent
{\large \bf Acknowledgements}

We acknowledge A. Perali, C. Castellani, and M. Grilli for helpful 
discussions.

\end{document}